\documentclass[convention,peer-reviewed]{aesconf}

\usepackage{amsmath}
\usepackage{amsfonts}
\usepackage{siunitx}
\usepackage{graphicx}
\usepackage{cite}
\usepackage{booktabs}
\usepackage{color}
\usepackage{url}
\usepackage{multirow}
\usepackage{hyperref}
\usepackage[capitalise,nameinlink]{cleveref}
\usepackage{graphicx}
\usepackage{grffile}
\usepackage{lipsum}
\usepackage{amssymb}

\DeclareMathAlphabet{\mathcal}{OMS}{cmsy}{m}{n}

\usepackage{titlesec}
\titlespacing*{\subsection}
{0pt}{0.1em}{0.1em}

\graphicspath{{./}{figures/}}

\usepackage{microtype}

\usepackage[numbers,square,comma, sort&compress]{natbib}

\title{{Efficient neural networks for real-time modeling of analog dynamic range compression}}

\author[]{Christian J. Steinmetz}
\author[]{Joshua D. Reiss}

\affil[]{Centre for Digital Music, Queen Mary University of London}

\correspondence{Christian J. Steinmetz}{c.j.steinmetz@qmul.ac.uk}

\lastnames{Steinmetz and Reiss}

\shorttitle{Neural compressor modeling}

\begin{document}

\twocolumn[
\vspace{0.3cm}

\maketitle % MANDATORY!
\vspace{0.2cm}
\begin{onecolabstract}
Deep learning approaches have demonstrated success in modeling analog audio effects. 
Nevertheless, challenges remain in modeling more complex effects that involve time-varying nonlinear elements, such as dynamic range compressors. 
Existing neural network approaches for modeling compression either ignore the device parameters, do not attain sufficient accuracy, or otherwise require large noncausal models prohibiting real-time operation. 
In this work, we propose a modification to temporal convolutional networks (TCNs) enabling greater efficiency without sacrificing performance. 
By utilizing very sparse convolutional kernels through rapidly growing dilations, our model attains a significant receptive field using fewer layers, reducing computation.
Through a detailed evaluation we demonstrate our efficient and causal approach achieves state-of-the-art performance in modeling the analog LA-2A, is capable of real-time operation on CPU, and only requires 10 minutes of training data.\vspace{1.0cm}
\end{onecolabstract}
]

\section{Introduction}

%Audio effects provide the ability to adjust perceptual attributes of audio signals such as loudness, timbre, pitch, spatialization, or rhythm, and form a core component of the tools used by audio engineers~\cite{wilmering2020history}.
While a significant amount of processing in audio and music production is performed digitally, there is a rich history of analog equipment that remains in high demand for its unique sonic signature.
As a result, there has been an interest in virtual analog modeling~\cite{karjalainen2006wave, yeh2009automated,  eichas2015block, eichas2017virtual, gerat2017virtual}, the task of constructing digital models to emulate these analog devices. 
While there are a range of traditional approaches in analog modeling, there has been a growing interest in neural network approaches~\cite{ wright2019real, damskagg2019distortion, martinez2020deep,chowdhury2020comparison}.
These approaches enable constructing emulations using only input-output measurements from the device, which has the potential to significantly lower the engineering effort in creating effect emulations.

Thus far, applications of neural networks for audio effect modeling have focused mostly on modeling vacuum-tube amplifiers~\cite{covert2013vacuum, schmitz2018nonlinear, zhang2018lstm, damskagg2019deep} and distortion circuits~\cite{damskagg2019distortion, wright2019real, ramirez2019modeling, chowdhury2020comparison, nercessian2021lightweight}. 
In contrast, time-varying nonlinear effects, like dynamic range compressors~\cite{giannoulis2012digital}, potentially pose a greater challenge in the modeling task due to their time-dependant nonlinearities, and have so far seen less attention.
A model of the 1176N compressor was proposed~\cite{martinez2020deep}, but it did not address the device control parameters and was evaluated only with electric guitar and bass signals. 
Modeling the LA-2A was addressed in~\cite{hawley2019signaltrain, mitchell2020exploring}, and while their model captured the overall characteristics of the device, it exhibits artifacts, is noncausal, and not capable of real-time operation, limiting its utility in audio engineering contexts.
Recently, temporal convolutional networks (TCNs) have shown success in modeling dynamic range compression~\cite{steinmetz2020msc, steinmetz2020mixing}, however, these models are also noncausal and computationally expensive. %and utilize a significant amount of training data. 

%Furthermore, these models have yet to be evaluated by listeners, leaving uncertainly about the degree to which they capture the characteristics of the original effect. 

To address these limitations we propose a more efficient and causal formulation of the TCN with the aim of facilitating real-time operation on CPU. 
We realize that while computation across the temporal dimension can be parallelized in the TCN, computations through the depth of the network are sequential. 
Therefore shallower networks provide one route towards greater efficiency, yet often at the cost of smaller context window sizes, also know as the receptive field, which may limit the accuracy of the model.

Our proposed efficient TCN employs rapidly growing dilation factors, effectively enforcing very sparse convolutional kernels, which facilitates shallow networks that achieve the same receptive field as deeper networks.
We carry out a range of experiments to validate our proposed architecture in the task of modeling the analog LA-2A compressor. 
We demonstrate that our proposed TCN architecture with fewer layers and sparse kernels performs competitively with larger noncausal formulations, producing strong results in a listening test, while also running in real-time on CPU.
Additionally, we examine the role of dataset size and find that only 10 minutes of training data is required. 
We provide audio examples, code, and pre-trained models online\footnote{\hspace{-0.0cm}\scalebox{0.93}{\url{https://csteinmetz1.github.io/tcn-audio-effects}}}. 

\section{Background}

%\begin{figure}[]
%    \centering
%    \includegraphics[width=\linewidth]{figs/audio-effect-task.drawio.pdf}
%    \caption{A neural network $g(x,\phi)$ is used to emulate an audio effect $f(x,\phi)$ with control parameters $\phi$.}
%    \label{fig:audio-effect}
%\end{figure}

We consider an audio effect $f(x, \phi)$ that takes as input an audio signal $x \in \mathcal{X}$ and a set of $P$ parameters $\phi \in \mathbb{R}^P$ that control the operation of the system, producing a corresponding processed version of the signal $y \in \mathcal{Y}$.
In the case of an analog effect, $x$ and $y$ are continuous time signals.
Since we aim to create a digital emulation, we utilize discrete time measurements, treating these signals as vectors $x,y \in \mathbb{R}^{S}$ with S samples.

Our aim is to construct a neural network $g_\theta(x, \phi)$ that produces a signal $\hat{y}$ perceptually indistinguishable from the output $y$ of the real effect. 
The modeling process involves training $g_\theta(x, \phi)$ with a dataset of $E$ examples $\mathcal{D} = \{(x_i, y_i, \phi_i) \}_{i=1}^{E}$ containing input-output recordings ($x_i, y_i$) at different device configurations $\phi_i$ . 
A loss function $\mathcal{L}(\hat{y}, y)$ is used to measure the difference between the output of the network and the target system, which provides a means to update the weights $\theta$ through a given number of optimization steps.
A successful model will accurately capture the behavior of the system across the space of control parameters $\Phi$ as well as the space of all possible input signals $\mathcal{X}$.

\begin{figure}[]
    \centering
    \includegraphics[width=\linewidth,trim={0.0cm 0.2cm 0.0cm 0.4cm},clip]{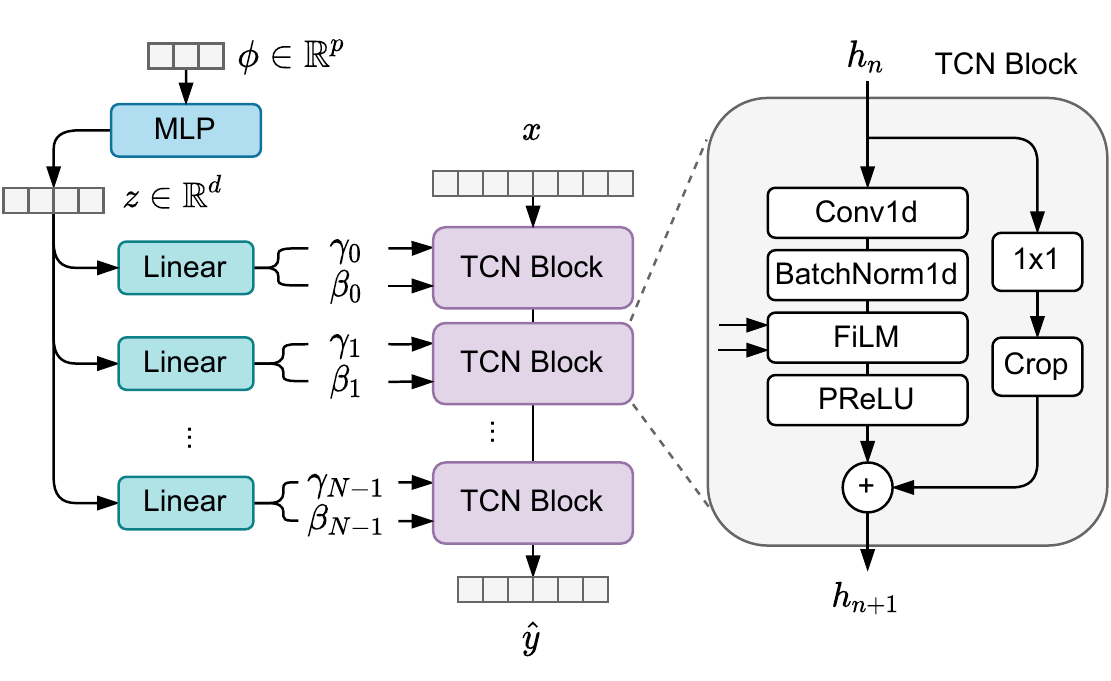}
    \vspace{-0.4cm}
    \caption{TCN~\cite{steinmetz2020mixing} with a series of convolutional blocks along with conditioning module (MLP) that adapts the gain $\gamma_n$ and bias $\beta_n$ at each layer as a function of the control parameters $\phi$.}
    \label{fig:tcn-arch}
    \vspace{-0.3cm}
\end{figure}

\subsection{Related work}

While there has been significant work in neural network approaches for distortion-based audio effects~\cite{covert2013vacuum, zhang2018lstm, schmitz2018nonlinear, damskagg2019deep, damskagg2019distortion, wright2019real, ramirez2019modeling, kuznetsov2020differentiable, wright2020perceptual, martinez2019general, martinez2020deep, chowdhury2020comparison, nercessian2021lightweight}, there has been less work in modeling analog dynamic range compression. 
The 1176N compressor was addressed in \cite{martinez2020deep}, where the authors utilized a range of different architectures including convolutional, recurrent, and a combination of the two. 
While their objective evaluation and listening test indicated strong performance in the task of modeling this compressor, their approach was limited in that they only considered one configuration of the device parameters. 
In addition, they trained and evaluated their models using only electric guitar and bass signals at a sample rate of 16\,kHz, potentially limiting application to other sources. 

A dataset containing measurements from the analog LA-2A was presented in \cite{hawley2019signaltrain}, as well as a model using an autoencoder operating on spectral representations. 
While their approach modeled the device parameters, used a wide range of content (voice, music, noises, etc.), and operated at 44.1\,kHz, it was found to exhibit noticeable artifacts. 
Further experimentation found reducing the diversity of sources in training and evaluation improved performance, but architectural modifications were not successful in addressing the artifacts~\cite{mitchell2020exploring}.

Similar to the feedforward WaveNet~\cite{rethage2018wavenet} employed in modeling the 1176N~\cite{martinez2020deep} and distortion effects~\cite{damskagg2019deep}, a modified TCN was proposed for modeling a range of effects including compression, along with the control parameters~\cite{steinmetz2020msc, steinmetz2020mixing}. 
Their approach replaced the gated convolution with feature-wise linear modulation (FiLM) to adapt based on the control parameters. 
This approach achieved state-of-the-art performance in modeling the LA-2A, however, these models are relatively large and noncausal, prohibiting real-time operation.

This architecture of the state-of-the-art TCN is shown in Fig.~\ref{fig:tcn-arch}. 
It consists of residual blocks, composed of 1-dimensional convolutions with increasing dilation factors, followed by batch normalization, conditional feature-wise linear modulation (FiLM)~\cite{perez2018film}, and a PReLU~\cite{he2015delving} nonlinearity. 
To obtain a large receptive field multiple blocks are stacked with a dilation factor that grows as a power of 2 as the depth of the network increases.
A network with $N$ layers uses convolutions with a dilation at layer $n \in 0, 1, ..., N-1$ given by $d_n = 2^{n}$.
This enables a larger receptive field in a more efficient manner using progressively more sparse convolutional kernels.

The FiLM operation enables adaptation of the network behavior based on the control parameters. 
This involves an affine transformation of intermediate activations $h_{n}$ with a set of scaling $\gamma_{n}$, and bias $ \beta_{n}$ parameters for each channel that are unique to each layer. 
This operation at the $n^{\textrm{th}}$ layer is given by $F(h_{n,c}, \gamma_{n,c}, \beta_{n,c}) = \gamma_{n,c} h_{n,c} + \beta_{n,c},$
where $c$ is the channel index. 
In order to generate the scaling and bias parameters for each layer, a multilayer perceptron (MLP) projects the device control parameters to an embedding $z \in \mathbb{R}^{d}$, shown in Fig.~\ref{fig:tcn-arch} Left. 
A linear layer at each block uniquely adapts $z$, the global conditioning, to produce $2C_n$ values, where $C_n$ is the number of convolutional channels at the $n^{\textrm{th}}$ layer. 
%Note that the $\gamma_{c}$ and $ \beta_{c}$ parameters will adapt at inference based on the device control parameters, extending the expressivity of the model. 
%This form of conditioning is not only simpler, but has demonstrated superior performance as a conditioning mechanism in convolutional networks in many domains, including audio \cite{meseguer2019conditioned, zeghidour2020wavesplit, petermann2020deep}.

\section{Proposed method}

In the design of a TCN for real-time operation we combine a causal formulation of the TCN along with an overall shallower network by using convolutions with rapidly growing dilation factors. 

We first consider the requirement for noncausality, which imparts a lower-bound on the latency our system can achieve. 
While noncausality may aid in the modeling task, a causal TCN \emph{should} be capable of modeling our causal analog system.
We propose to do so by adopting causal convolutions, which are a common feature of TCNs~\cite{bai2018empirical}, and have been utilized in previous work on modeling distortion effects~\cite{damskagg2019deep}. 

In the case of the noncausal TCN~\cite{steinmetz2020mixing}, the input receptive field is split evenly between the past and future samples, such that a delay of $\approx$150\,ms is required for adequate ``look-ahead''. 
% (we find it cannot, as discussed in Sec.~\ref{sec:runtime}).
%This noncausal formulation is illustrated in Fig.~\ref{fig:padding}a, where no padding is applied, the output activations will be shorter in length then the input, and the output field is centered. 
%\begin{figure}
%    \centering
%    \includegraphics[width=\linewidth]{figs/padding.drawio.pdf}
%    \caption{Padding approaches for an input signal of $N$ samples with a convolutional kernel of size 3. a) Noncausal TCN formulation where no padding is applied~\cite{steinmetz2020mixing}, b) Standard ``same'' padding, which is also noncausal, c) Causal padding.}
%    \label{fig:padding}
%\end{figure}
To achieve causality, the output must be a function only of current and previous inputs, which can be achieved with adequate padding.
Causal convolutions pad the input on the left with $r - 1$ samples, where $r$ is the size of the receptive field of the model. 
This ensures that the output at each time-step is a function only of the current and past inputs.
%Note that padding equally on the left and right to ensure the output is of the same size will not be causal (often called ``same'' padding), as will no padding (often called ``valid'' padding). 

Since we opt to only pad the input signal, and not the intermediate activations, the output of each convolution will be smaller than the input. 
This requires we crop the residual connections in each TCN block.
%The residual connections are implemented with 1x1 grouped convolutions, which enables only scaling of the features. 
Care must be taken to perform cropping of the residual connections correctly depending on the causality of the model.
In the noncausal case, a central crop is taken across the temporal dimension, while in the causal case, this crop selects the last $S$ samples, where $S$ is the number of time-steps at the output of the convolution.

%This padding is illustrated in Fig.~\ref{fig:padding}c for a convolution with a kernel size of $3$.

\begin{figure}[]
    \centering
    \vspace{0.1cm}
    \includegraphics[width=\linewidth,trim={1.5cm 0.4cm 0.8cm 0.3cm},clip]{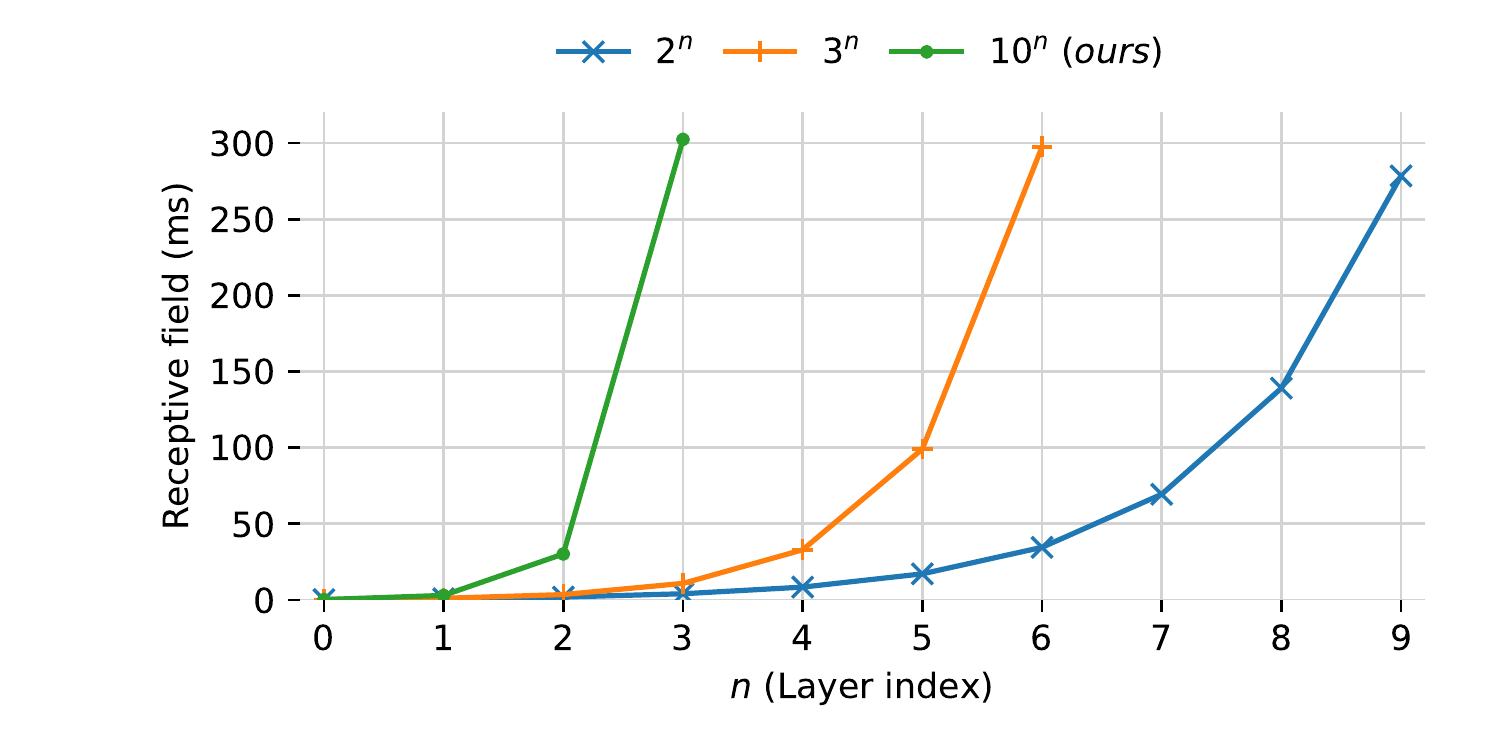}
    \vspace{-0.2cm}
    \caption{Effect of the dilation growth on the receptive field of the TCN in milliseconds at $f_s=44.1$\,kHz as a function of the number of layers. Here we use a TCN with kernel size $K=13$.}
    \label{fig:receptive-field}
    \vspace{-0.0cm}
\end{figure}

However, causality does not necessarily produce a model capable of real-time operation. 
The model must additionally be able to process a buffer of $S$ samples in less than $S / f_s$ seconds, where $f_s$ is the sample rate. 
To reduce the computational complexity of the TCN, we acknowledge that while computation across the temporal dimension can be parallelized within the frame, computation through the depth of the network cannot. 
%This imposes a run-time constraint that is a function of the model depth, as well as other factors like the number of convolutional channels, the hardware platform (CPU/GPU), and the convolution implementation~\cite{chowdhury2021rtneural}.
Therefore, one straightforward route to decreasing the run-time involves simply constructing shallower networks. 
Unfortunately, this often comes at the cost of a smaller receptive field assuming the kernel size and dilation pattern are maintained, which often leads to a decrease in the model accuracy. 

To rectify this, we propose a simple modification. 
We can achieve a comparable receptive field with fewer layers by using dilation factors that grow more rapidly in comparison to the base 2 convention~\cite{yu2016dilated}, $d_n = 2^{n}$ with $n \in 0, 1, ..., N-1$.
In the case of the TCN with $N$ layers, the receptive field at the $n^{\textrm{th}}$ layer is given by the recursion $r_{n} = r_{n-1} + (K - 1) \cdot d$, where $K$ is the kernel size, and $d$ is the dilation factor. 
The receptive field at the first layer is given by the kernel size $r_0 = K$.
Following this, we plot the receptive field of different TCNs as a function of the network depth $N$ and the dilation growth with kernel size $K=13$ in Fig.~\ref{fig:receptive-field}.

The noncausal TCN from~\cite{steinmetz2020mixing} requires $N=10$ layers in order to achieve a receptive field of approximately 300\,ms. 
While less common, some recent speech synthesis models employ more aggressive dilation patterns, such as $d_n = 3^{n}$ \cite{yang2020multi, tian2020tfgan}.
However, this still requires $N=7$ layers to achieve a comparable receptive field. 
Therefore, we propose to use an even larger dilation growth, $d_n = 10^{n}$, which enables only $N=4$ layers to achieve the same receptive field as previous methods.
Since the dilation factors are progressively increased, the first few layers still use relatively dense filters, as in the previous approaches, yet with the later layers achieving much larger receptive field.
%Using such large dilation factors reduces the overall model expressivity by enforcing the convolutional kernels to be very sparse.
To our knowledge, there has been no investigation of models that utilize dilation factor growth at this rate.

\section{Experimental design} \label{sec:experiments}

\subsection{Dataset}

To validate our proposed TCN in the analog audio effect modeling task we consider the SignalTrain dataset\footnote{(Version 1.1) \url{https://zenodo.org/record/3824876}}~\cite{hawley2019signaltrain}. 
This dataset provides approximately 20 hours of input-output recordings at $f_s = 44.1$\,kHz from the analog LA-2A dynamic range compressor.
%\footnote{\url{https://www.uaudio.com/blog/la-2a-analog-obsession}}.
It covers a diverse range of audio content including individual instruments, loops, and complete musical pieces, in addition to tones and noise bursts.
This compressor features two control parameters: a binary switch that places the device in either \emph{compress} or \emph{limit} mode, as well as a continuous peak reduction parameter that controls the amount of compression as a function of the input level. 
The dataset provides audio processed by the compressor at 40 different parameter configurations, enabiling the ability to model the device at multiple different configurations.
We use the same training, validation, and test split in the original dataset.

\subsection{Models}
We re-implement the TCN from~\cite{steinmetz2020mixing, steinmetz2020auraloss}, which we denote TCN-324-N. This model has 10 layers with a dilation pattern given by $d_n = 2^{n}$, where each layer includes 32 channels. 
This model is noncausal and achieves a receptive field of 324\,ms at $f_s =$ 44.1\,kHz. 
We also adapt the LSTM architecture proposed in \cite{wright2019real}, which we denote LSTM-32, since it features a single recurrent layer with 32 hidden units. 
We consider variants of the TCN to investigate the impact of noncausality, and the ability to achieve greater efficiency with shallower networks and larger dilation factors. 
These also employ 32 channels, but utilize a more rapidly growing dilation pattern given by $d_n = 10^{n}$, enabling the use of fewer layers with a similar receptive field. 
%Our goal is to develop a network that accurately models the device, but is also capable of real-time operation. 

In order to observe the impact of the receptive field on model performance, we train variants of the efficient TCNs with receptive fields of 101\,ms (TCN-100), 302\,ms (TCN-300), and 1008\,ms (TCN-1000). 
To observe the need for noncausality, we train each model in both causal and noncausal formulations. Models ending in ``-N'' are noncausal, while those ending in ``-C'' are causal.  
We also investigate the amount of training data required.
We train the TCN-300-C model with subsets of the dataset that contain only 10\% and 1\% of the training data by splitting the training set by the parameter configurations, and randomly sampling an equal amount of audio from each of these configurations. 

\setlength{\tabcolsep}{6.0pt}
\renewcommand{\arraystretch}{0.9}
\begin{table*}[!htbp]
    \centering
    \vspace{-0.0cm}
    \resizebox{\linewidth}{!}{
    \begin{tabular}{l c c c c c c c c c c}
        \toprule
         \textbf{Model} & $K$ & $N$ & $d$ & $C$ & $P$ & \textbf{R.f.} & \textbf{RT (CPU/GPU)} & \textbf{MAE} $\downarrow$ & \textbf{STFT} $\downarrow$ & \textbf{LUFS} $\downarrow$ \\ 
        \midrule
        TCN-324-N~\cite{steinmetz2020mixing} & 15 & 10 & 2 & 32 & 162\,k  & 324 ms &  0.5x / 17.1x &  1.70e-2 & 0.587 & 0.520 \\ \midrule
        TCN-100-N & 5 & 4 & 10 & 32 & 26\,k & 101\,ms & 4.2x / 37.1x & 1.58e-2 & 0.768 & 1.155 \\
        TCN-300-N & 13 & 4 & 10 & 32 & 51\,k & 302\,ms & 1.8x / 37.3x & \textbf{7.66e-3} & 0.600 & 0.602 \\
        TCN-1000-N & 5 & 5 & 10 & 32 & 33\,k & 1008\,ms & 0.5x / 26.4x & 1.20e-1 & 0.736 & 0.934 \\
        %TCN-900-N & 15 & 30 & 2 & 512\,k & \,G & 900\,ms & 0.0x / 0.0x & 6.65e-3 & 0.505 & 0.455 \\
        \midrule
        TCN-100-C & 5 & 4 & 10 & 32 & 26\,k & 101\,ms & 5.0x / 37.2x & 1.92e-2 & 0.770 & 1.225 \\
        TCN-300-C & 13 & 4 & 10 & 32 & 51\,k & 302\,ms & 2.2x / 37.3x & 1.44e-2 & 0.603 & 0.761 \\ 
        %TCN-370-C & 5 & 3 & 60 & 19\,k & 0.70\,G & 377 ms & 3.1x / 43.1x &  1.12e-1 & 0.936 & 1.313 \\ 
        TCN-1000-C & 5 & 5   & 10 & 32 & 33\,k & 1008\,ms & 0.6x / 26.4x & 1.17e-1 & 0.692 & 0.899 \\
        \midrule
        LSTM-32 & - & - & - & - & 5\,k & - &  0.9x / 2.8x &  1.10e-1 & \textbf{0.551} & \textbf{0.361} \\
        \bottomrule \\
    \end{tabular}}
    \vspace{-0.3cm}
    \caption{Performance on the LA-2A test set. Models ending with -N are noncausal, and those ending -C are causal. $K$ is the kernel size, $N$ is the number of layers, $d$ is the dilation growth factor, $C$ is the number of convolutional channels, and $P$ is the total number of trainable parameters. R.f is the receptive field in milliseconds. The real-time factor (RT) is reported on CPU and GPU with a frame size of 2048 samples.} %Further details on these measurements are provided in Sec. \ref{sec:runtime}.}
    \vspace{-0.0cm}
    \label{tab:model-comparision}
\end{table*}

\subsection{Training}
All models were trained with a batch size of 32 and inputs of 65536 samples ($\approx$1.5\,s at 44.1\,kHz) for a total of 60 epochs on a single GPU. 
The only augmentation applied during training was a phase inversion of the input and target signals applied with probability $p=0.5$~\cite{hawley2019signaltrain}. 
We employed Adam~\cite{kingma2014adam} with an initial learning rate of $3 \cdot 10^{-4}$, decreasing the learning rate by a factor of 10 after the validation loss had not improved for 10 epochs.
In evaluation, we used the model weights from each configuration that achieved the lowest validation loss during training.
Additionally, we used automatic mixed precision to decrease training time and memory consumption, which we found had negligible effect on the model performance or training stability. 
We have made the code to reproduce these experiments available online\footnote{\url{https://github.com/csteinmetz1/micro-tcn}}.

\subsection{Loss function}

For training we used a combination of the error in the time and frequency domains.
We compute the mean absolute error (MAE) for the time domain component $\mathcal{L}_\textrm{time}$ and the multi-resolution short-time Fourier Transform error~\cite{yamamoto2019probability, steinmetz2020auraloss} for the frequency domain $\mathcal{L}_\textrm{freq}$ component as used in previous work~\cite{steinmetz2020mixing}.
The overall loss function is given as a sum of these two terms $
    \mathcal{L}_\textrm{overall} = \mathcal{L}_\textrm{time} + \alpha \cdot \mathcal{L}_\textrm{freq}.$
We used $\alpha=1$ in all experiments.
In effect this weights the frequency domain loss more greatly, due to the differing scales of the terms. 
%which we found produced results that align better with perception. 

\setlength{\tabcolsep}{3pt}
\begin{table}[t]
    %\centering
    \resizebox{\linewidth}{!}{
    \begin{tabular}{l c c c c c c c c c}
        \toprule
        \textbf{Model} & $C$ & $P$ & \textbf{RT} & \textbf{MAE} & \textbf{STFT} & \textbf{LUFS} \\ 
        \midrule
        324-N & 32 & 162\,k & 0.5x / 17.1x & 1.70e-2 & \textbf{0.587} & \textbf{0.520}  \\
        324-N & 16 & 47\,k &  1.3x / 17.1x & 4.38e-2 & 0.796 & 1.305 \\
        324-N* & 8 & 16\,k &  2.2x / 17.1x & 5.29e-2 & 1.143 & 1.315 \\ \midrule
        300-C & 32 & 51\,k &  2.2x / 33.4x & \textbf{1.44e-2} & 0.603 & 0.761  \\
        \bottomrule \\
    \end{tabular}}
        \vspace{-0.4cm}
    \caption{TCN-324 models using fewer convolutional channels. $^*$Model diverged during training.}
    \vspace{-0.1cm}
    \label{tab:param-comparision}
\end{table}

\subsection{Metrics}

We considered three metrics for the objective evaluation of the models. 
The first two are components of the training objective, the MAE of the time-domain signal, and the multi-resolution STFT error (denoted STFT). 
As a perceptually informed metric, we define the loudness error as the absolute error between the loudness of the prediction and target signals computed using the ITU-R BS.1770 perceptual loudness recommendation~\cite{bs1770, steinmetz2021pyloudnorm}.
With this metric we can measure to what degree the perceived loudness was captured by the model, which is likely correlated with the application of the correct gain reduction as a result of compression.

\section{Results} \label{sec:result}

Results comparing our causal and efficient TCNs to previous approaches are shown in Table~\ref{tab:model-comparision}. 
The model hyperparameters, $K$ kernel size, $N$ number of layers, and $d$ dilation growth factor are reported, along with the number of model parameters $P$ and the receptive field in milliseconds. %achieved at $f_s = 44.1$\,kHz.
We report the real-time factor (RT) for a frame size of 2048 samples, which is described in more detail in Sec.~\ref{sec:runtime}.
%Models ending in ``-N'' are noncausal, while those ending in ``-C'' are causal. 
These results suggest that causal formulations of the TCN are able to achieve comparable performance to their noncausal variants, with the most significant difference being that noncausal models appear to achieve slightly superior time domain performance and lower dB LUFS error.
However, the TCN-1000-C model is an exception, performing slighter better than the TCN-1000-N across all metrics.
%It also appears as if improved time-domain performance (lower MAE) requires a model with a greater number of parameters comparatively. 
%This effect is less prominent in the STFT error, but the correlation appears as well.

With regards to the TCNs, it appears that models with around 300\,ms of receptive field achieve superior performance. 
Although, this may be due to the smaller number of parameters in the models with different receptive field.
Nevertheless, the \mbox{TCN-1000-C} model, which features few parameters and the largest receptive field, is not capable of real-time operation. 
Our efficient TCNs, which employ very large dilation growth factors and are shallower than the TCN-324-N model, yet have comparable receptive field and performance while using a third of the parameters and providing up to four times faster run-time on CPU. 

%While the TCN-300-C model achieves comparable receptive field to the TCN-324-N model, it features fewer parameters, and therefore a faster run-time. 

Notably, the LSTM-32 model achieves the best performance across both the STFT and LUFS metrics, but an order of magnitude worse with respect to the time-domain performance (MAE).
However, it is difficult to make a conclusion based soley on these objective metrics, which motivates our listening study as outlined in Sec~\ref{sec:listening}
The strong performance of the LSTM-32 demonstrates the major advantage of recurrent models, namely that they are able to achieve an adaptive receptive field in a parameter efficient manner. 
In this case, the LSTM-32 uses 32x fewer parameters than the TCN-324 model.  
Nevertheless, while this class of models is parameter efficient, processing across the temporal dimension cannot be parallelized.
In this case, the LSTM-32 model is not capable of real-time operation on CPU in the PyTorch implementation even when compiled via torchScript\footnote{\url{https://pytorch.org/docs/stable/jit.html}}.
Additionally, the LSTM-32 model required over 8 times longer to train~(108\,hr) compared to the TCN-300-C model~(13\,hr).

\subsection{Parameter scaling}

To further demonstrate the efficacy of larger dilation factors, we demonstrated that merely scaling down the parameters of the TCN-324-N model does not provide comparable accuracy and efficiency.
We trained narrower variants of the TCN-324-N model with fewer convolutional channels, as shown in Table~\ref{tab:param-comparision}. 
We found that while scaling down the width of these models does increase the real-time factor, it comes at the cost of performance, with the TCN-300-C significantly outperforming these variants. 
This strengthens our claim that using very sparse convolutional kernels is an effective method for achieving sufficient receptive field without sacrificing performance in the modeling task.

\subsection{Data efficiency} \label{sec:data}
\setlength{\tabcolsep}{3pt}
\begin{table}[]
    \centering
    \resizebox{\linewidth}{!}{
    \begin{tabular}{l r c r c c c c c c}
        \toprule
        \textbf{Model} & \textbf{Data} & \textbf{Config} & \textbf{Total} & \textbf{MAE} & \textbf{STFT} & \textbf{LUFS}\\ 
        \midrule
        324-N & 100\% & 30 m & 19.5 h & 1.70e-2 & 0.587 & \textbf{0.520} \\
        \midrule
        300-C & 100\% & 30 m & 19.5 h & 1.44e-2 & 0.603 & 0.761  \\
        300-C & 10\%  & 3.0 m & 1.9 h & \textbf{1.38e-2} &\textbf{0.587} & 0.630 \\
        300-C & 1\%   & 0.3 m & 11.3 m & 1.40e-2 & 0.599 & 0.740 \\
        \bottomrule \\
    \end{tabular}}
    \vspace{-0.4cm}
    \caption{TCN-300-C with varying amount of data.}
    %\vspace{-0.3cm}
    \label{tab:data-comparision}
\end{table}
While the SignalTrain dataset provides 20 hours of recordings from the LA-2A, we investigated the requirement for such a large dataset. 
We split the original training dataset into random subsets, with a balanced number of examples for each parameter configuration. 
The 10\% subset contains a total of 1.9 hours of audio with 3 minutes of audio per configuration of the compressor parameters. 
Furthermore, the 1\% subset results in a total of just 11 minutes of audio in total, with only 18 seconds per configuration. 
Results for the TCN-300-C trained with these subsets are compared against TCNs trained with the complete dataset in Table~\ref{tab:data-comparision}.

We found reducing the size of the training dataset did not significantly impact performance.
Surprisingly, there is an improvement in performance using the smaller training subsets.
We hypothesize this could be due to some special characteristics of the random subsets that were selected.
For example, perhaps more samples with tones and noise bursts were selected, which could be more informative, or vice versa. 
This indicates that modeling related analog dynamic range compression effects could be achieved with significantly smaller datasets.
This greatly lowers the burden in creating such datasets and agrees with findings from previous works in modeling other effects~\cite{wright2019real, schmitz2018nonlinear}. 
%\vspace{-0.05cm}

\subsection{Compute efficiency} \label{sec:runtime}

\begin{figure}
    \centering
    \includegraphics[width=\linewidth,trim={0.4cm 0.2cm 0.45cm 0.4cm},clip]{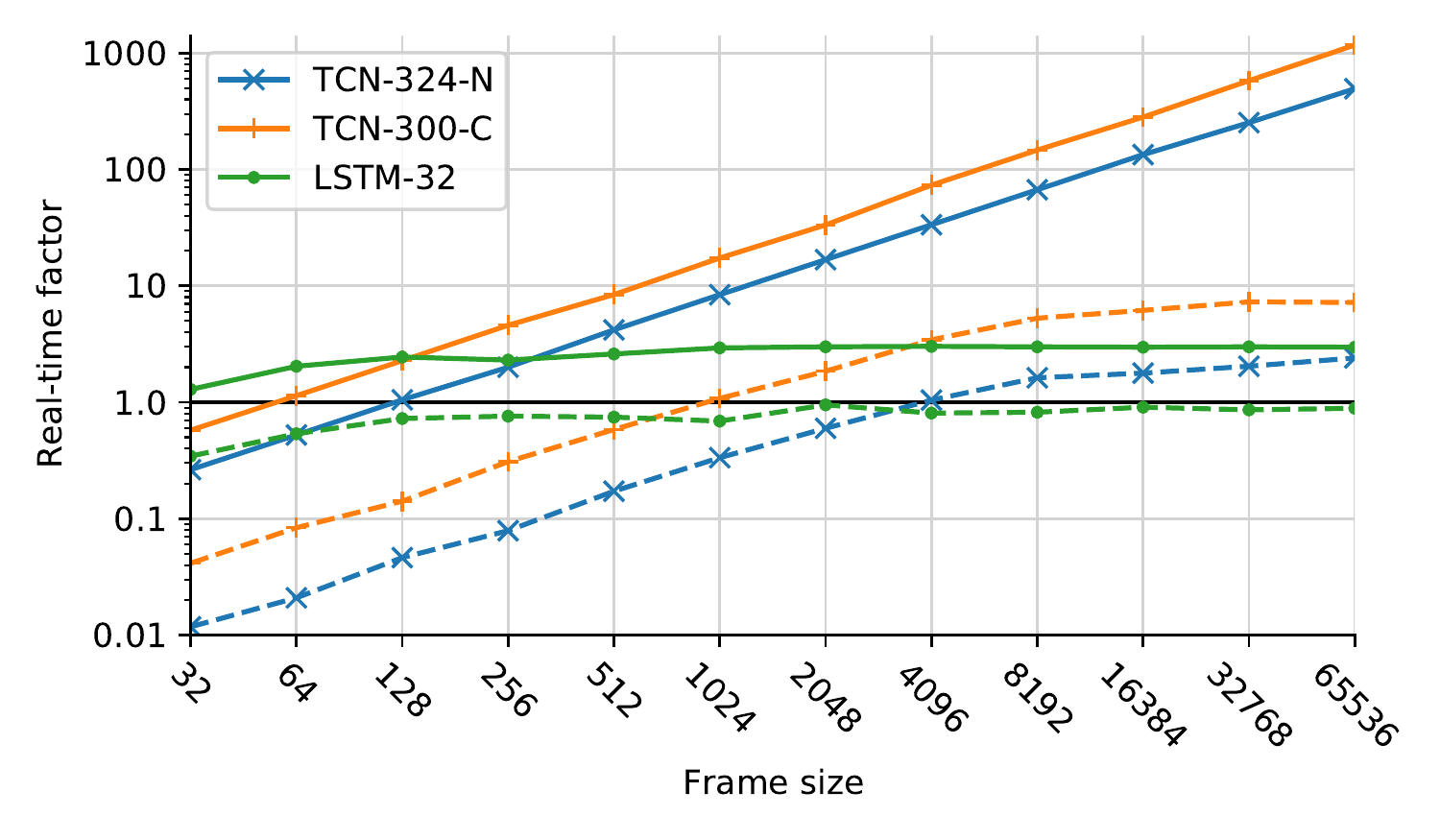}
    %\vspace{-0.7cm}
    \caption{RT on GPU (solid) and CPU (dashed) at different frame sizes. RT greater than 1 is required for real-time operation.}
    %\vspace{-0.3cm}
    \label{fig:real-time}
\end{figure}
We investigated the run-time of these models in a block-based implementation that aims to mimic a standard audio effect. 
The real-time factor (RT) is defined as 
\begin{equation}
\text{RT} := \frac{S}{T \cdot f_s},
\end{equation}
where $S$ is the number of samples processed at a sampling rate of $f_s$, and $T$ is the time in seconds to process those $S$ samples. 
We measure the real-time factor at power of 2 frame sizes, $F \in {32, 64, ..., 65536}$, on both GPU and CPU. 
For GPU, measurements are performed on a RTX 3090, and for CPU, measurements are performed on a 2018 MacBook~Pro with an Intel Core i7-8850H @ 2.6 GHz. 
Results are shown in Fig.~\ref{fig:real-time} on GPU (solid lines) and CPU (dashed lines). 
In this block-based formulation, the TCN models require a buffer of past samples such that we pass an input of $S + r - 1$ samples, where $S$ is the number of output samples and $r$ is the receptive field in samples.

For the LSTM, the real-time factor on both GPU and CPU is constant with respect to the frame size, which is due to the inability to parallelize computations across the temporal dimension. 
In our PyTorch implementation, we found the LSTM was close, but not able to achieve real-time operation. 
On the other hand, we found the real-time factor for the TCN model is proportional to the frame size, with larger frame sizes producing greater real-time factors as a result of greater parallelization, both on CPU and GPU.
This enables real-time operation on CPU at frame sizes down to 1024 samples, which we found also to be the case in our implementation of the model in a JUCE plugin.
%These results indicate that on GPU, the TCN-300-C model achieves real-time operation with frame sizes down to 64 samples, and can take advantage of a $1000 \times$ speedup on longer sequences, unlike the LSTM. 
%Although, it seems that at frame sizes larger than 8192 samples this improvement begins to level off. 

This understanding of recurrent and convolutional models can help guide the architecture design process for modeling effects. 
In cases where very low latency is required, assuming a recurrent model of sufficient size can run in real-time on the target platform, these models provide a good option. 
On the other hand, convolutional models demonstrate a clear advantage in that larger frame sizes will provide a significant speedup, useful in offline use cases, such as rendering a mixdown, or when using neural audio effects in other contexts, such as automatic mixing~\cite{steinmetz2020mixing}. 
These results represent a worse-case scenario, since optimized C++ implementations may achieve a speedup compared to the PyTorch models used in our analysis~\cite{wright2019real, damskagg2019distortion, chowdhury2020comparison, chowdhury2021rtneural}.

\subsection{Listening study}\label{sec:listening}

%While objective metrics enable comparison of relative performance, they provide little insight into the degree to which a model emulates the device from a perceptual standpoint.
To further evaluate model performance, we carried out a multistimulus listening test, similar to MUSHRA~\cite{mushra}. 
Five passages from the test set were used, each around 12 seconds in duration. 
We processed these stimuli using the SignalTrain model, the LSTM-32 model, and our proposed causal TCN-300-C model trained with 1\% of the dataset ($\approx 10$\,min). 
We did not include a low quality anchor as there was no clear choice in the case of dynamic range compression~\cite{maddams2012autonomous, ma2015intelligent}.
We used webMUSHRA \cite{schoeffler2018webmushra}, which enabled the study to be performed online, and allowed participants to instantaneously switch between different stimuli in order to facilitate comparison of small differences. 

\begin{figure}[t]
    \centering
    \vspace{0.2cm}
    \includegraphics[width=\linewidth,trim={1.0cm 0.1cm 0.7cm 0.75cm},clip]{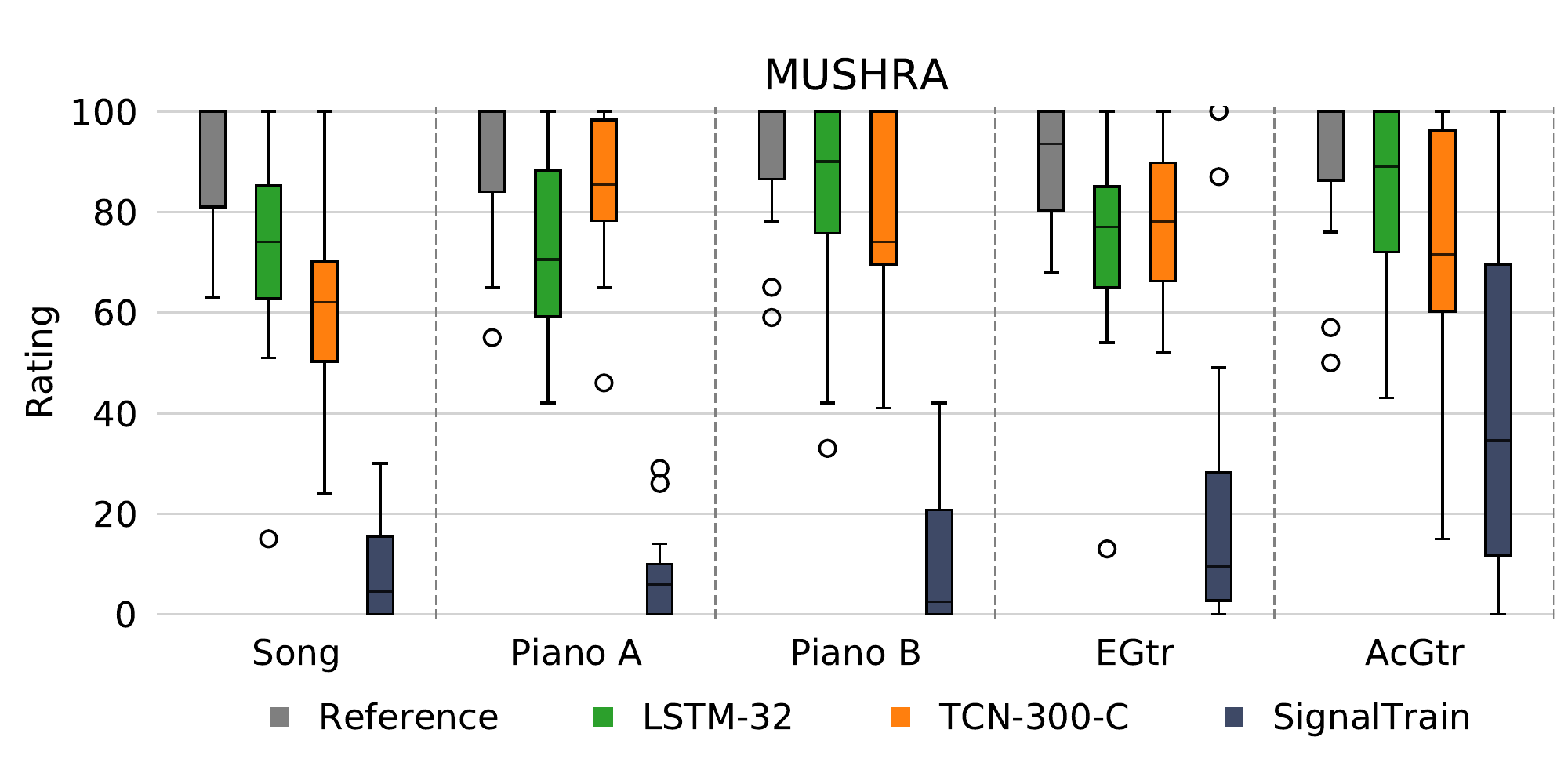}
    %\vspace{-0.3cm}
    \caption{Ratings of the five passages from the MUSHRA style listening studying with 18 participants after the post-screening process.} %The TCN-300-C model used in the evaluation was trained with only 1\% of the training dataset.}
    \label{fig:mushra}
    %\vspace{-0.4cm}
\end{figure}

We enlisted 19 participants, all of whom reported experience with audio engineering and were familiar with the LA-2A. 
We performed a post-screening analysis to assess the participants, and removed ratings from one participant who assigned the reference a score of less than 50 in 4 of the 5 passages.%, leaving ratings from  18 participants. 
Results from the remaining 18 participants are presented in Fig.~\ref{fig:mushra}. 

%During the study, listeners are presented with a reference signal, the original output from the analog LA-2A. 
%They are then tasked with providing a rating from 0 to 100 for four different conditions.
%Three of these conditions are the models we are studying, as well as the reference.

Both the LSTM-32 and TCN-300-C performed slightly below the reference. 
With some stimuli the median rating of the LSTM-32 is greater (Song, Piano B, AcGtr), while at other times the TCN-300-C is greater (Piano A, EGtr).
In contrast, it is clear that participants noticed the strong noise-like artifacts produced by the SignalTrain model.
Some participants struggled to differentiate between the reference and LSTM-32 and TCN-300-C models, as they rated the reference lower than these models in some cases. 
This is evident from the high variance in the ratings for the reference. 

To formalize these observations, we performed the Kruskal-Wallis $H$-test, which indicated a difference in the median rating of the models ($F = 186.7, p = 3.21\cdot10^{-40}$). 
A post hoc analysis using Conover's test of multiple comparisons revealed a significant difference in the ratings for the reference and the LSTM-32 ($p_{\text{adj}} = 3.65\cdot10^{-11} $) and TCN-300-C ($p_{\text{adj}} = 6.84\cdot10^{-9}$). 
This indicated, that while challenging, listeners likely perceived a small difference among the models in comparison to the reference.

Nevertheless, it appears there is no significant difference in the median ratings between the LSTM-32 and TCN-300-C ($p_{\text{adj}} = 0.37 $).
%Additionally, we find there is a significant difference between the LSTM-32 ($p_{\text{adj}} = 1.78e\cdot10^{-25} $) and TCN-300-C ($p_{\text{adj}} = 8.95\cdot10^{-29} $) in comparison to the previous SignalTrain model.
These results appear to agree with comments from participants, where both the LSTM-32 and TCN-300-C models were found to very closely capture the character of the LA-2A without imparting artifacts, but differ in cases of strong gain reduction, letting some transients pass through more so than the analog LA-2A. 

% not sure if this conversation is need?
%\section{Discussion}
%TCNs are well-suited for audio effect modeling since no downsampling in time occurs throughout the network. 
%This contrasts with autoencoding approaches, like SignalTrain, that must accurately reconstruct high frequency information in the decoder, which can be challenging. 
%While sample-based LSTMs also perform no downsampling, and provide comparable accuracy, they are an order of magnitude slower. 
%Additionally, they do not provide significant speedup on GPU, like the parallelizable TCNs.
%The most significant caveat of the TCN is that it must employ a sufficiently large receptive field.
%This may not be feasible for some effects with low frequency oscillators, such as phaser or chorus effects~\cite{wright2020lfo, martinez2019general}.

%Our investigations only serve to demonstrate that it is possible for this formulation of the TCN models to achieve sufficient accuracy in the modeling task while also achieving real-time operation. 
%Further investigation is needed to determine which kinds of audio signals are most informative, and the required density of the parameter sampling to achieve sufficient performance. 
%Future work could consider more advanced techniques to designing efficient neural networks such as quantization, distillation, and pruning, along with optimized implementations for target hardware, i.e. general purpose CPUs. 
\newpage
\section{Conclusion}

We demonstrated that TCNs employing causal convolutions with rapidly growing dilation factors enable shallow networks to achieve sufficient receptive field in a compute-efficient manner.
This causal and efficient TCN formulation was effective in modeling the analog LA-2A dynamic range compressor, ultimately enabling real-time operation on CPU.
A listening study found that our proposed model achieved a high level of perceptual similarity to the original device, outperforming the previous SignalTrain model, using only 1\% of the full training dataset in the process. 
However, our results indicated that while challenging, listeners were often able to differentiate the emulations from the original device, leaving room for further improvement.
Directions for future investigation involve optimizations in platform specific implementations for further efficiency in real-time operation, as well as investigating how TCNs with rapidly growing dilation factors generalize to other audio effects and related audio signal processing tasks. 

\section*{Acknowledgements}
This work is supported by the EPSRC UKRI Centre for
Doctoral Training in Artificial Intelligence and Music
(EP/S022694/1).

\bibliographystyle{jaes}

% Reference to bibliography file.

\bibliography{refs}

\end{document}